\documentclass[11pt, oneside]{article}   	
\usepackage[T1]{fontenc}
\usepackage{lmodern}
\usepackage{geometry}                		
\usepackage{graphicx}				
\usepackage{amsmath}		
\usepackage{amssymb,amsfonts,amsthm}
\usepackage{mathtools,cool}
\usepackage{braket}

\usepackage[cal=boondoxo]{mathalfa}

\title{\textbf{\Large A Note on Classical Aspects of the Four Dimensional Anomaly-Free Twistor String}}
\author{Christian Kunz \\ \small{\textit{E-mail:} \href{mailto:kunz.christian.321@gmail.com}{kunz.christian.321@gmail.com}}}
\usepackage{plain}
\usepackage{hyperref}

\newcommand{\ud}{\mathrm{d}}
\numberwithin{equation}{section}
\allowdisplaybreaks

\begin{document}
  \maketitle
  \tableofcontents
  
\begin{abstract}
  The recently introduced anomaly-free twistor string in four dimensions is shown to be defined not just in flat, but also in curved twistor space. Further, arguments are given that the classical limit of the corresponding string field theory, if it exists, is related to  general relativity, in particular to the Isenberg and Yasskin construction using teleparallel gravity. For spacetimes of Petrov type D with two shear-free null congruences the construction can be simplified using two-dimensional twistor manifolds.
\end{abstract}
\pagebreak

\section{Introduction}
     In \cite{Kunz:2020, Kunz_1:2020} a four dimensional anomaly-free twistor string has been presented. This twistor string reproduces in the Neveu-Schwarz (NS) sector the familiar N$^k$MHV amplitudes of N=8 supergravity \cite{Skinner:2013,Geyer:2014} and the Einstein-Yang-Mills (EYM) amplitudes of \cite{Adamo:2015}. The one-loop amplitudes in both the NS and the Ramond (R) sector are modular invariant and exhibit proper unitary factorization properties.\\
     
     This note aims to shed more light on the spacetime interpretation of the twistor string. Inspired by the fact that the 10-dimensional ambitwistor string can be defined on curved spacetime (with modifications) \cite{Adamo:2014, Adamo:2018}, here it is shown that the twistor string works classically on deformed twistor space, even without modifications. Further, assuming that the classical limit of a string field theory of the twistor string (e.g. as proposed in \cite{Kunz_1:2020}) exists, the question arises whether there is a correspondence with general relativity. We give arguments that indeed there is a possibility to establish a correspondence along the lines of the ambitwistor program proposed by Isenberg and Yasskin \cite{Yasskin:1982, Isenberg:1986, Isenberg:1987}. And for conformal spacetimes of Petrov type D with two shear-free two-fold principal null congruences (SFR-PND), Araneda \cite{Araneda:2019} investigated a correspondence with two-dimensional twistor manifolds which maps nicely to our twistor string model, simplifying the general construction.\\
     
     In section 2 we argue that the target space of the twistor string, due to the presence of two twistors or 'bi-twistors', is naturally parametrized by spacetime, extended to a superspace with anticommuting coordinates, and that the deformation of the derivative on the worldsheet due to curvature can be included into the Lagrange multipliers of gauge symmetries. Therefore, the twistor string action does not need to be modified for a deformed twistor space.
     
     In section 3 we exploit the fact that a basis for the matter field portion of Hilbert states of the twistor string splits naturally into two halves. Each half can be used to create a correspondence with self-dual and anti-self-dual spacetimes, respectively, resulting in a bidual geometry similar to the one proposed by Isenberg and Yasskin \cite{Isenberg:1987}. Because of the gauging of the conformal translations this leads naturally to a teleparallel gravity \cite{Krssak:2018} and hopefully to an embedded teleparallel equivalent of general relativity (TEGR). These arguments are not to be taken as a proof of the correspondence. Rather, they serve as an incentive for further investigation.
     
     In section 4 we use Araneda's results on two-dimensional twistor manifolds \cite{Araneda:2019} to argue that for conformal spacetimes of Petrov type D with two SFR-PND's we can identify two two-dimensional twistor manifolds in twistor space and another two such manifolds in dual twistor space, on which string fields can be defined. Such spaces include Kerr-(A)dS black holes and might be conducive for further investigation of twistor string properties.
     
     The last section contains summary and discussion.
     
\section{Curved Twistor Space}
  \label{Curved}
     
   The action of the four dimensional anomaly-free twistor string (left-moving and closed) is \cite{Kunz:2020}
    \begin{multline}
   S_0 = \frac{1}{2 \pi} \int\mathop{}\negthickspace{\ud^2\mathop{}\negthickspace z} \left\{
   \frac{1}{2} \sum_{i=1}^2 \left( Y_i\overline\partial{Z_i} - Z_i \overline\partial{Y_i} + \Theta_i\overline\partial{\Psi_i}  + \Psi_i\overline\partial{\Theta_i} \right)
   \right.\\
  + \sum_{i,j=1}^2 \lambda_i a_{1 i j} \tilde{\phi}_j
  + \sum_{i,j=1}^2 \tilde{\lambda}_i a_{2 i j} \phi_j
  + \sum_{i,j=1}^2\tilde{\lambda}_i b_{i j} \lambda_j
  + \left(Y_1 Y_2 \right) \mathop{}\negthickspace \vec{c} \, \cdotp \vec{\tau}  \mathop{}\negthickspace \begin{pmatrix} Z_1 \\ Z_2 \end{pmatrix}
  \left. \vphantom{\sum_i} \right\},
  \label{action}
  \end{multline}
  where for $i=1,2$ $Z_i$ are twistors, $Y_i$ dual twistors, $\Psi_i$ fermionic bi-spinors, and $\Theta_i$ fermionic dual bi-spinors, with components:
 \begin{equation*}
 Z_i = \binom{\lambda_{i \alpha_i}}{\mu_i^{\dot{\alpha}_i}}, Y_i = \binom{\tilde{\mu}_i^{\alpha_i}}{\tilde{\lambda}_{i \dot{\alpha}_i}},
 \Psi_i = \binom{\phi_{i \alpha_i}}{\psi_i^{\dot{\alpha}_i}}, \Theta_i = \binom{\tilde{\psi}_i^{\alpha_i}}{\tilde{\phi}_{i \dot{\alpha}_i}}.
 \end{equation*}
 
  The action has an SU(2) gauge symmetry between the two twistors and additionally bosonic gauge symmetries of the conformal translations and fermionic gauge symmetries for conformal supertranslations (see the next paragraph for a justification of using the word 'supertranslation').\\
 
  Compared to other twistor strings in the literature, one peculiarity of this twistor string is the presence of bi-twistors. On the other hand, in order to specify a spacetime point uniquely, two twistors ($\sim$ light rays) are needed \cite{Penrose:1986}. This way, deformed twistor space can be regarded as a local twistor bundle over spacetime. Because of the fermionic gauge symmetries the model is better defined on an extended spacetime, a superspace $(x^{\alpha \dot{\alpha}}, \theta_{1ij}^{\alpha \dot{\alpha}}, \theta_{2ij}^{\dot{\alpha} \alpha})$ with a total of 32 anticommuting coordinates $\theta$. The bosonic translation gauge symmetries correspond to translations in $x^{\alpha \dot{\alpha}}$ with Lagrange multiplier $b_{ij}$ and the fermionic supertranslation gauge symmetries correspond to translations in $\theta_{1ij}^{\alpha \dot{\alpha}}$ and $\theta_{2ij}^{\dot{\alpha} \alpha}$ with Lagrange multipliers $a_{1ij}$ and $a_{2ij}$, respectively. We have incidence relations\footnote{Notice that in contrast to the usual ambitwistor in four dimensions \cite{Geyer:2014} there are no incidence relations for the $\phi_i$ and $\tilde{\phi}_i$ fields.}
  \begin{align}
  &\mu_i^{\dot{\alpha}} = x^{\alpha \dot{\alpha}} \lambda_{i \alpha} + \theta_{2ij}^{\dot{\alpha} \alpha} \phi_{j \alpha},
  &\psi_i^{\dot{\alpha}} &= \theta_{1ij}^{\alpha \dot{\alpha}} \lambda_{j \alpha},\nonumber\\
  &\tilde{\mu}_i^{\alpha} = - x^{\alpha \dot{\alpha}} \tilde{\lambda}_{i \dot{\alpha}} + \theta_{1ij}^{\alpha \dot{\alpha}} \tilde{\phi}_{j \dot{\alpha}},
  &\tilde{\psi}_i^{\alpha} &= - \theta_{2ij}^{\dot{\alpha} \alpha} \tilde{\lambda}_{j \dot{\alpha}}.
  \label{incidence}
  \end{align} 
  
  Fermionic gauge transformations \cite{Kunz:2020} can be used to reduce the incidence relations for $\mu_i^{\dot{\alpha}}$ and $\tilde{\mu}_i^{\alpha}$ in \eqref{incidence} to the simpler ordinary ones, $\mu_i^{\dot{\alpha}} = x^{\alpha \dot{\alpha}} \lambda_{i \alpha}$ and $\tilde{\mu}_i^{\alpha} = - x^{\alpha \dot{\alpha}} \tilde{\lambda}_{i \dot{\alpha}}$. Then tangent vectors to worldsheets do not contain odd components. For such tangent vectors local twistor transport on the worldsheet becomes \cite{Penrose:1986}:
 \begin{align*}
 &\overline\nabla Z_i = \overline\partial Z_i \!+\! \binom{P_{\alpha \beta \dot{\alpha} \dot{\beta}} \overline\partial x^{\beta \dot{\beta}} \mu_i^{\dot{\alpha}}}{\overline\partial x^{\alpha \dot{\alpha}} \lambda_{i \alpha}},
 &\Theta_i \overline\nabla \Psi_i = \Theta_i \overline\partial \Psi_i \!-\! \binom{\tilde{\lambda}_{i \dot{\alpha}} (\overline\partial x^{\beta \dot{\alpha}} \lambda_{i \beta} \!-\!  x^{\gamma \dot{\alpha}} P_{\gamma \beta \dot{\gamma} \dot{\beta}} \overline\partial x^{\beta \dot{\beta}} \mu_i^{\dot{\gamma}})}{- \tilde{\phi}_{i \dot{\alpha}} \theta_{1ij}^{\gamma \dot{\alpha}} P_{\gamma \beta \dot{\gamma} \dot{\beta}} \overline\partial x^{\beta \dot{\beta}} \mu_j^{\dot{\gamma}}},\\
 &\overline\nabla Y_i = \overline\partial Y_i \!-\! \binom{\overline\partial x^{\alpha \dot{\alpha}} \tilde{\lambda}_{i \dot{\alpha}}}{P_{\alpha \beta \dot{\alpha} \dot{\beta}} \overline\partial x^{\beta \dot{\beta}} \tilde{\mu}_i^{\alpha}}, 
 &\Psi_i \overline\nabla \Theta_i = \Psi_i \overline\partial \Theta_i \!-\!  \binom{- \phi_{i \alpha} \theta_{2ij}^{\dot{\gamma} \alpha} P_{\gamma \beta \dot{\gamma} \dot{\beta}} \overline\partial x^{\beta \dot{\beta}} \tilde{\mu}_j^{\gamma}}{\lambda_{i \alpha} (\overline\partial x^{\alpha \dot{\beta}} \tilde{\lambda}_{i \dot{\beta}} \!+\!  x^{\alpha \dot{\gamma}}P_{\gamma \beta \dot{\gamma} \dot{\beta}} \overline\partial x^{\beta \dot{\beta}} \tilde{\mu}_i^{\gamma})}.
  \end{align*} 
 Here $P_{\alpha \beta \dot{\alpha} \dot{\beta}}$ is the Schouten tensor (vanishing in the vacuum) and the second column follows from the first column and incidence relations \eqref{incidence}. When replacing all $\overline{\partial}$ in the action \eqref{action} by $\overline{\nabla}$, inserting above expressions for $\overline{\nabla}$, and using the incidence relations \eqref{incidence}, all extra terms cancel against each other, in self-fulfilling manner, and we are left with the original action. But more strikingly, when just considering any individual term in the action containing a single $\overline{\nabla}$ the difference between $\overline\nabla$ and $\overline\partial$ can be added to the Lagrange multiplier fields in the action \eqref{action}, this time by using the reduced incidence relations. The same result even holds in the general case where tangent vectors are not limited, extra terms in the covariant derivative for $\lambda_i, \mu_i, \phi_i,$ and $\psi_i$ are some linear combinations of the other three spinors, likewise for the dual spinors, and reduced incidence relations are inserted.\\
 
  Therefore, the twistor string action as given in \cite{Kunz:2020} is well defined on curved twistor space. The incidence relations and gauge symmetries have been critical to establish this fact. It should be noted that quantization on curved target spaces has not been considered here.

\section{Spacetime Correspondence}
  \label{Correspondence}
 
  In appendix A of \cite{Kunz_1:2020} it was shown that in the R sector physical states naturally split into two different (dual) Hilbert spaces based on which matter field zero modes annihilate the vacuum. Assuming that a string field theory exists, we can consider two subsets of string fields, each restricted to one particular set of matter field zero modes. In the NS sector two similar subsets can be considered, one excluding the dual twistor field $-\frac{1}{2}$ modes and the other one excluding the twistor field $-\frac{1}{2}$ modes. In each subset we now restrict the string fields to be on-shell, but without the requirement that they are annihilated by the gauge currents $F_{i j},  G_{1i j}$, and $G_{2i j}$ which belong to the (super)conformal translation symmetries. Then the string fields, after gauging of the bi-twistors, look like arising from twistor string actions considered in \cite{Adamo:2012, Adamo_3:2013} for self-dual and anti-self-dual gravity and supergravity, especially in the R sector \footnote{Our twistor string model has an NS sector because the matter fields are worldsheet spinors, not just scalars, but this does not prevent the correspondence.}, and we can directly apply the results of \cite{Adamo:2012, Adamo_3:2013} to claim that our twistor string includes a correspondence between twistor space and spacetime as spelled out in \cite{Isenberg:1986}, namely that the twistor string lives in a deformed bitwistor\footnote{Here the word bitwistor of Isenberg \& Yasskin \cite{Isenberg:1986} refers to the product of twistor space with its dual, not to the two twistors in our model. To make the distinction, we will use the word 'bi-twistor' with a hyphen to refer to the two twistors in the string.} space ($\mathcal{P}^{\alpha \beta}, \mu_L, \mu_R$) ($\mu_L, \mu_R$ being vertical 2-forms) in a one-to-one correspondence with a bidual geometry ($\mathcal{M}^8, T\mathcal{M} \!=\! T_L\mathcal{M} \!\oplus\! T_R\mathcal{M}, g \!=\! g_L \!\oplus\! g_R, \nabla \!=\! \nabla_l \!\oplus\! \nabla_R$) which is a deformation of an 8-dimensional complex Minkowski space and  kind of a product of self-dual and anti-self-dual 4-dimensional geometries with metrics $\!g_{\!L \atop{\!R}}\!$ and connections $\!\nabla_{\!L \atop{\!R}}\!$ on distributions $\!T_{\!L \atop{\!R}}\!\mathcal{M}$.\\
  
  Now what happens to the correspondence when we do not disregard any gauge symmetries? In the twistor space this amounts, of course, to using the anomaly-free twistor string fully in the proper way. In spacetime this means restricting considerably the set of fields belonging to physical particle states, breaking the conformal symmetry, and resulting in a  teleparallel gravity \cite{Krssak:2018} with, of course, no guarantee that it will turn out to be an embeddable TEGR. If this guarantee could be established, we would have a solution to a variant of the Isenberg \& Yasskin ambitwistor program. The fact that in the NS sector the twistor string leads to the correct gravitational and EYM scattering amplitudes \cite{Kunz_1:2020} is a promising sign. On the other hand, the scattering amplitudes in the R sector do not look as expected and, therefore, the interpretation of this part of the model remains an open question. We will add a couple of comments in the last section \ref{Discussion} concerning this issue. Concerning spacetime supersymmetry, although the spectrum in the R sector \cite{Kunz_1:2020} has resemblance to conformal supergravity, the one in the NS sector \cite{Kunz_1:2020} seems to indicate there is no overall conventional spacetime supersymmetry.

\section{Petrov Type D Spacetimes with Two Principal Shear-Free Null Congruences}
  \label{PetrovD}
  Araneda \cite{Araneda:2019} investigated Petrov type D spacetimes with two SFR-PND's and showed that the two shear-free PND's are associated with two independent two-dimensional twistor manifolds representing moduli spaces for $\beta$-surfaces (can also been seen as consequence of the twistor version of the Kerr theorem \cite{Penrose:1986}), and that the conformal manifold structure of the foliations by $\beta$-surfaces induces an almost-complex structure on the spacetime compatible with a conformal metric. The same can, of course, be done independently in conjugate manner for dual two-dimensional twistor manifolds using $\alpha$-surfaces.\\
  
  Now, as shown in appendix A of \cite{Kunz_1:2020}, the physical states of the twistor string depend only on the two-dimensional spinor parts $\lambda_{i \alpha}$ and $\tilde{\lambda}_{i \dot{\alpha}}$ of the bi-twistors. Therefore, we can generate the same subsets of string fields as in the previous section, but now restricted to the two-dimensional twistor manifolds as moduli spaces, on-shell covariantly constant on the associated $\beta$- or $\alpha$-surfaces, one for each member of the bi-twistor. This leads again to a correspondence between a bitwistor space of two pairs of 2-dimensional twistor manifolds and an 8-dimensional bidual geometry of left and right (not necessarily self-dual or anti-self-dual) conformal Petrov type D spacetimes with two SFR-PND's. Again, there is no proof that the eventual embedding of the 4-dimensional spacetime is possible and results in a gravity equivalent to one that satisfies the Einstein or similar field equations.

\section{Summary and Discussion}
  \label{Discussion}
  In this note we looked at some classical aspects of the four-dimensional anomaly-free twistor string. We showed that because of the incidence relations and gauge symmetries the string is classically well defined on deformed twistor space. Then we argued that the classical limit of a string field theory is related to general relativity, based on the Isenberg \& Yasskin program using teleparallel gravity. For Petrov type D spacetimes with two SFR-PND's the correspondence between spacetime and twistor space simplifies considerably. No proofs have been given that the eventual embedding of the spacetime into the bidual geometry works and results in a theory equivalent or at least similar to general relativity. The hope is that the arguments given here would spur deeper investigations.\\
  
  One interesting aspect of the twistor string model is that it has graviton-like excitations in both the NS and the R sector \cite{Kunz_1:2020}. The graviton in the NS sector \footnote{There is another graviton-like excitation in the NS sector \cite{Kunz_1:2020} containing mixed twistor and dual twistor field $-\frac{1}{2}$ modes with no obvious interpretation and being disregarded here.} has been shown to lead to expected scattering amplitudes, but this is not true for the spin 2 excitations in the R sector. If the Isenberg \& Yasskin construction of section \ref{Correspondence} succeeds it remains to be seen whether the twistor string leads to a bi-metric gravity model in spacetime \cite{DeFelice:2020} (and references therein, see also \cite{Hossenfelder:2008} for an unusual approach) or something else. A peculiarity of the model is that all spin $\frac{1}{2}$ particles are in the R sector and, therefore, couple directly to the spin 2 excitation in the R sector but only indirectly to the graviton in the NS sector.\\
  
  Concerning the construction in the previous section \ref{PetrovD} one upside resides in the fact that the two-dimensional twistor manifolds can be determined explicitly for interesting spacetimes like the Kerr black hole and used as background for further investigation of the twistor string properties. This seems more straightforward than trying to put the twistor string on reduced non-Hausdorff twistor spaces \cite{Fletcher:1988, Fletcher:1989} with only implicit geometric connection to spacetime.

\bibliography{TwistorString}
\end{document}